\documentclass{raa}   

\usepackage{graphicx,times}             
\usepackage{natbib}

\usepackage{amssymb,amsmath}
\bibpunct{(}{)}{;}{a}{}{,}

\usepackage[pagebackref=true]{hyperref}

\begin{document}

\title{The influence of the Sun and Moon on the observation of very high energy gamma-ray sources using EAS arrays}

   \volnopage{Vol.0 (20xx) No.0, 000--000}      
   \setcounter{page}{1}          

   \author{Tao Wen 
      \inst{1,2}
   \and Songzhan Chen
      \inst{2,3}
   \and Benzhong Dai
      \inst{1}
   }

   \institute{School of Physics and Astronomy, Yunnan University, 650091 Kunming, Yunnan, China; {\it bzhdai@ynu.edu.cn, wentao@ihep.ac.cn}\\
        \and
             Key Laboratory of Particle Astrophyics, Institute of High Energy Physics, Chinese Academy of Sciences, 100049 Beijing, China;{\it chensz@ihep.ac.cn}\\
        \and
             Tianfu Cosmic Ray Research Center, 610000 Chengdu, Sichuan,  China\\
\vs\no
   {\small Received 20xx month day; accepted 20xx month day}}

\abstract{With great advance of ground-based extensive air shower array, such as LHAASO and HAWC, many very high energy (VHE) gamma-ray sources have been discovered and are been monitored regardless of the day and the night. Hence, the Sun and Moon would have some compact on the observation of gamma-ray sources, which have not been taken into account in previous analysis. In this paper, the influence of the Sun and Moon on the observation of very high energy gamma-ray sources when they are near the line of sight of the Sun or Moon is estimated. The tracks of all the known VHE sources are scanned and several VHE sources are found to be very close to the line of sight of the Sun or Moon during some period. The absorption of  very high energy gamma-ray by sunlight  is estimated with detailed method  and some usefully conclusions are achieved.   The main influence is the block of the Sun and Moon on gamma-ray and their shadow on the cosmic ray background. The influence is investigated considering the detector angular resolution and some strategy on data analysis are proposed to avoid the underestimation of the gamma-ray emission. 
}

   \authorrunning{T.Wen, S.Z.Chen ,B.Zh.Dai }            
   \titlerunning{Sun and Moon's Influence on Very High Energy Gamma-Ray Source Observation }  

   \maketitle

\section{Introduction} \label{sec:intro}
Very high energy (VHE, E$>$0.1 TeV) gamma-rays, located at the highest energy band of the cosmic electromagnetic radiation, are a powerful probe for astrophysics and fundamental physics under extreme conditions. Thanks to the advancements in ground-based gamma-ray detectors, our knowledge about the VHE gamma-ray universe has made impressive progress over the past two decades. The successful operation of the second generation Imaging Atmospheric Cherenkov Telescopes (IACTs), such as H.E.S.S.~\citep{Jung2001}, MAGIC~\citep{BAIXERAS2003247}, and VERITAS~\citep{2002APh....17..221W}, has significantly increased the number of detected VHE gamma-ray sources from about 10 to over 200 since 2004. Due to the limitation of this technique which adopt optical detectors, the observation can only be implemented during the clear and moonless night. Therefore, the influence of the Sun and Moon and their light on the very high energy gamma-ray sources are ignored for a long time.

Recently, the sensitivity of another ground-based detector technique EAS arrays has greatly improved thanks to the successful operation of the new generation arrays, such as HAWC and LHAASO. The HAWC \citep{2012NIMPA.692...72D,HAWC} and LHAASO \citep{He2018} observatories has detected 65 and 90 VHE sources\citep{2020ApJ...905...76A,2023arXiv230517030C}, respectively. Since the EAS arrays adopt particle detectors, they have a large field of view which cover 1/7 of the whole sky at each moment, and they are usually operated with nearly full duty cycle regardless of the day and the night. In such a situation, some of the gamma-ray sources may be very close to the line of sight of the Sun or Moon during observation which would lead to the underestimation of the sources flux. Therefore, the influence of the Sun and the Moon on the very high energy gamma-ray sources near the line of sight should be carefully studied and some strategy is needed to guide the data analysis to avoid the underestimation of the gamma-ray emission.

When a gamma-ray sources is very close to the line of sight of the Sun or the Moon, the measurement of the gamma-ray emission may be affected due to four factors. Firstly, when a source is blocked by the Sun or the Moon, the gamma-rays from the source will not reach the earth.
Secondly, when a source is close to the line of sight of the Sun, the gamma-rays from the source will suffer the $\gamma-\gamma$ absorption by the sunlight during their propagation to the earth. The quantitative absorption should depend on the energy of the gamma-rays and also depend on the space angle between source and the Sun. The extinction of the VHE  gamma-ray  by sunlight has been noticed in a brief paper \citep{2022RNAAS...6..148L} when studying the TeV gamma-ray background, however, no detailed results are presented and the quantitative  impact on gamma-rays from a specific source is not estimated.   Thirdly, the Sun and Moon will produce a shadow on the cosmic ray background, which will lead to a deficit source that cancel the gamma-ray signals.  These shadows have been well measured by  EAS arrays with high significance and were  adopted to calibrate detectors and perform physics measurements \citep{1998PhRvD..59a2003A,2003APh....20..145A,2011ApJ...729..113A,2011PhRvD..84b2003B,2017APh....90...20B,2021PhRvD.103d2005A}. Fourthly, the Sun could also produce gamma-ray emission due to interaction of cosmic ray with the solar atmosphere. The solar gamma-rays  have been well detected and studied using  Fermi-LAT data at GeV band \citep{2011ApJ...734..116A,2016PhRvD..94b3004N,2018PhRvL.121m1103L,2018PhRvD..98f3019T}. The emission at VHE band have been studied by several literature  \citep{2018PhRvD..98l3011A,2019ApJ...872..143B,CPC10.1088} and a positive signal was also reported by HAWC collaboration recently  \citep{2023PhRvL.131e1201A}.

In this paper, we will study the influence of the Sun and Moon on the very high energy gamma-ray sources near the line of sight. Especially, the absorption of  very high energy gamma-ray by sunlight will quantitative calculated for the first time. Then, we will explore some strategy on data analysis to avoid affection of the Sun and Moon on the measurement of the gamma-ray emission from the sources.  The paper is organized as follows: the known VHE gamma-ray sources that may be very close to the Sun and Moon are explored in Section 2.  The absorption of very high energy gamma-ray by sunlight is calculated in Section 3 including method and results. Some strategy to avoid the affection of the Sun and Moon are presented in Section 4.  Section 5 is the summary.

\section{VHE gamma-ray sources that could be close to the line of sight of the Sun and Moon} \label{sec:2}
\begin{figure}
\begin{center}
\includegraphics[width=16cm]{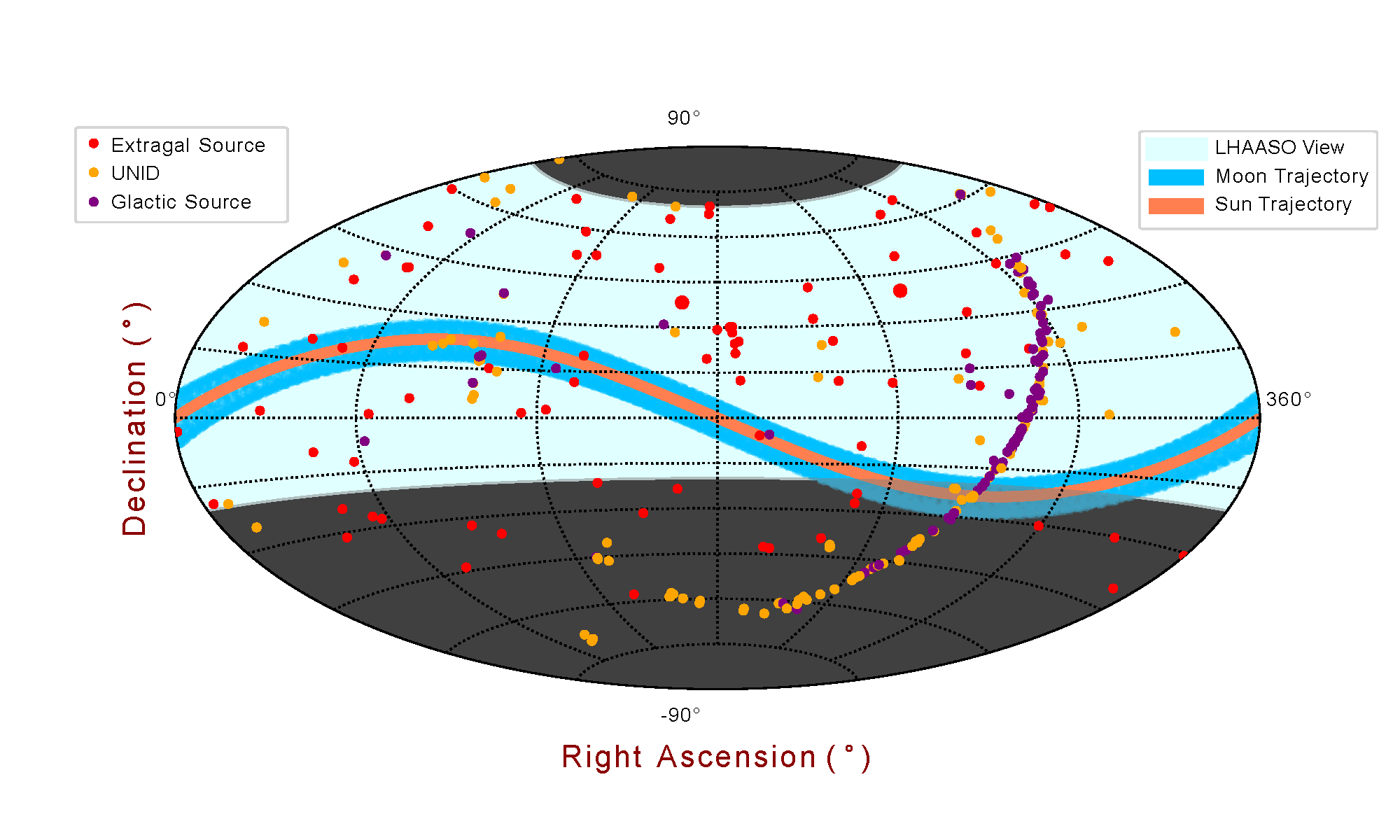}
\caption{\label{fig1} The track of Sun and Moon in the  Equatorial Coordinates. The Points are TeV Sources listed in  TeVCat, where the red points are Extragal Sources, the purple points are Galactic Sources, the orange points represent the UNID Sources. The  light cyan region is the field of view  of LHAASO. The orange line indicates the trajectory of Sun in one year. The skyblue belt is the trajectory of moon in 18.6 years.  }
\end{center}
\end{figure}

Due to the Earth's orbit around the Sun, the position of the Sun in the celestial coordinates varies annually. Figure \ref{fig1} shows the track of the Sun in the in celestial coordinates. The declination of the Sun can vary from -23.6$^{\circ}$ to 23.6$^{\circ}$. The track is about same every year. Due to the Moon’s orbit around the Earth, the position of the Moon in the celestial coordinates varies with a duration of 29.53 days. Besides this monthly variation, the varying range of the declination also vary with a duration of 18.61 year. The maximum (minimum) declination can change from 18.2$^{\circ}$ (-18.2$^{\circ}$) to 28.7$^{\circ}$ (-28.7$^{\circ}$). The track of the Moon in the celestial coordinates during the 18.61 year is also shows in Figure \ref{fig1}.  It is worth to note that both the Sun and Moon are extended source in the sky with a radius about 0.26$^{\circ}$.

Up to now, about 300 VHE gamma-ray sources have been detected according to TeVCat \footnote{\url{http://tevcat.uchicago.edu/}}. 
All these sources are also presented in the Figure \ref{fig1}.
According to the Figure \ref{fig1}, it is clear that some sources can be blocked by the Sun or Moon and some sources would be very close to the Sun or Moon during some period. Table \ref{tab1} list the 20 known VHE sources that with the closest space angle less than 2$^{\circ}$ from the Sun or Moon. The right ascension (R.A.) and declination (Dec) of the sources and their closest space angle to Sun and Moon are presented in the table \ref{tab1}. These sources include both Galactic and extra-galactic sources.  Obviously, these sources should be with careful treatment when measure their gamma-ray emission using the EAS array data. The impact of the   Sun and the Moon will be studied later.

\begin{table}
\begin{center}
\caption[]{ The  VHE sources  with the closest space angle less than $2^\circ$ from the Sun or Moon}\label{tab1}

 \begin{tabular}{l|cccc}
  \hline\noalign{\smallskip}
Source name &  Ra($^\circ$)      & Dec($^\circ$) & closest angle to Sun ($^\circ$) & closest angle to Moon ($^\circ$)  \\
  \hline\noalign{\smallskip}
HAWC J0543+233 & 85.78 & 23.40 & 0.313 & 0.325 \\
W 28 & 270.43 & -23.33 & 0.315 & 0.375 \\
3C 279 & 194.05 & -5.79 & 0.513 & 0.428 \\
HESS J1800-240B & 270.11 & -24.04 & 0.601 & 0.573 \\
RBS 0413 & 49.95 & 18.76 & 0.603 & 0.410 \\
HESS J1800-240A & 270.49 & -23.96 & 0.634 & 0.301 \\
HESS J1800-240C & 269.71 & -24.05 & 0.709 & 0.436 \\
IC 443 & 94.21 & 22.50 & 0.983 & 0.354 \\
Crab Pulsar & 83.63 & 22.01 & 1.307 & 0.558 \\
Crab & 83.63 & 22.01 & 1.308 & 0.560 \\
Terzan 5 & 266.95 & -24.81 & 1.413 & 0.563 \\
HESS J1804-216 & 271.13 & -21.70 & 1.735 & 0.426 \\
SNR G004.8+6.2 & 263.35 & -21.57 & 1.737 & 0.077 \\
2HWC J1309-054 & 197.31 & -5.49 & 1.765 & 0.278 \\
Kepler's SNR & 262.67 & -21.49 & 1.790 & 0.478 \\
VER J0521+211 & 80.44 & 21.21 & 1.928 & 0.241 \\
1ES 0647+250 & 102.69 & 25.05 & 2.157 & 0.559 \\
OJ 287 & 133.70 & 20.10 & 2.600 & 0.443 \\
GRB 180720B & 0.53 & -2.94 & 2.911 & 0.442 \\
HESS J1808-204 & 272.16 & -20.43 & 2.997 & 0.265 \\              \\
  \noalign{\smallskip}\hline
\end{tabular}
\end{center}
\end{table}

\section{The absorption of  VHE gamma-rays by sunlight} \label{sec:3e}
It is known that that VHE gamma-rays emitted from distant astronomical sources would be absorbed by low energy background photons through photon-photon interaction. These absorption  leads to opacity of the universe to the VHE gamma-rays. The absorption of gamma-rays at a specific energy $E$ is mainly due to the low energy photons with wavelength around $\lambda \sim$ 1.5($E$/1TeV) $\mu m$. The gamma-ray opacity above 100 TeV is mainly due to the absorption f Cosmic Microwave Background (CMB) \citep{2001ARA&A..39..249H}. The gamma-ray opacity bellow 100 TeV is mainly due to the absorption of Extragalactic background light (EBL) in the intergalactic space \citep{2010ApJ...712..238F,2021MNRAS.507.5144S}. The interstellar radiation fields (ISRF) within our Galaxy would also absorb the gamma-rays bellow 100 TeV with a visible fraction for some cases \citep{2006ApJ...640L.155M}. Before a cosmic VHE gamma-ray photon reaches the Earth, it should also pass through the  sunlight. Therefore, the gamma-ray should suffer the $\gamma-\gamma$ absorption by the sunlight. The density of sunlight dependent on the distance to the Sun $r$, which is in inverse proportion to $R_{\bigodot}^2$. The absorption should dependent on the space angle between a gamma-ray source and the Sun. In the following, the Crab Nebula will be taken as an example to estimate the absorption by sunlight.

\subsection{Estimation method}
\begin{figure}
\begin{center}
\includegraphics[width=16cm,height=10cm]{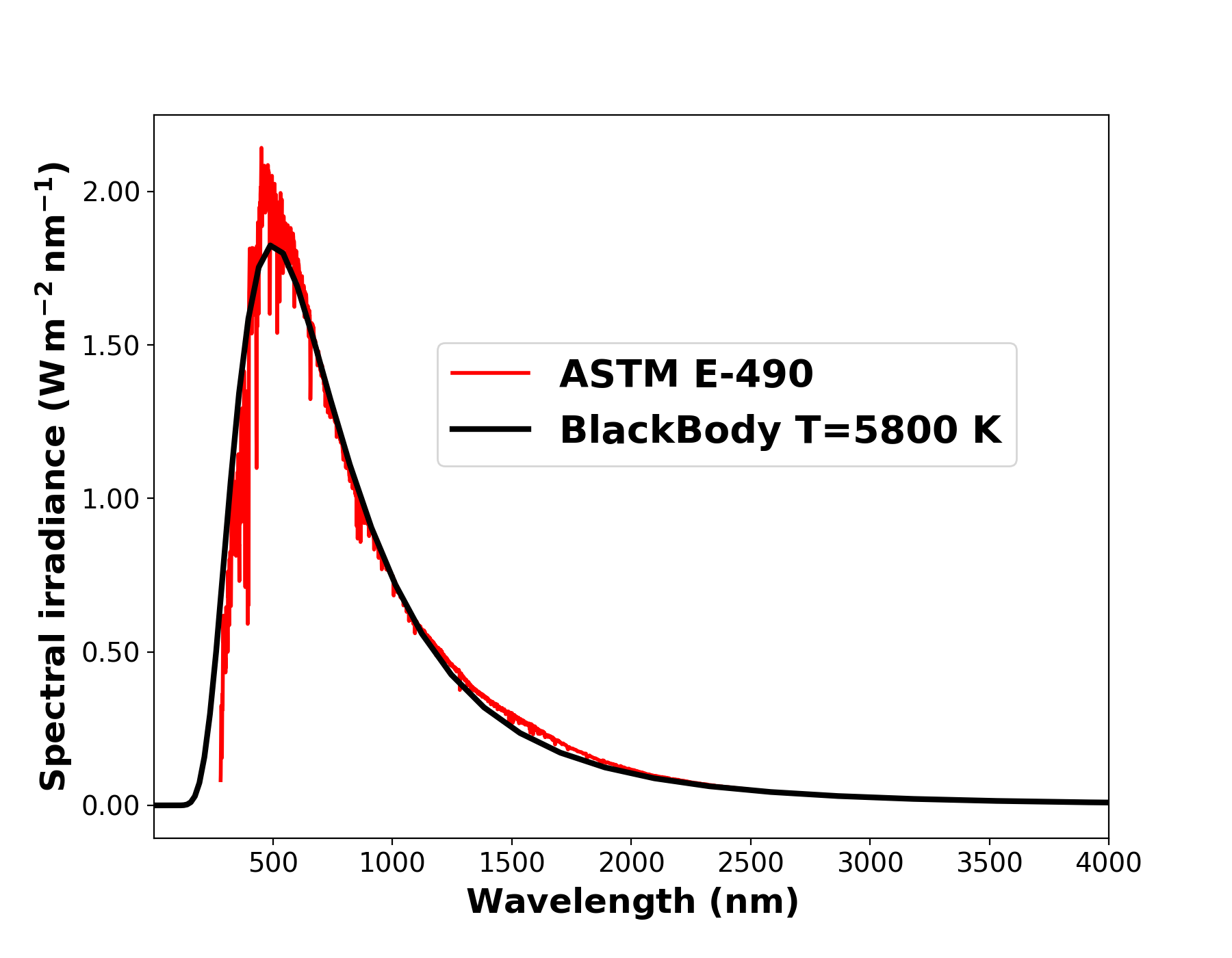}
\caption{\label{fig2} The luminosity of sunlight as the function of photon wavelength. The black line is the expected luminosity at Earth from a black-body spectrum with a temperature of 5800K at the surface of the  Sun. The red line indicates the measurement achieved by American Society for Testing and Materials (ASTM).}
\end{center}
\end{figure}

To quantity estimate the absorption by sunlight, the radiation spectrum from Sun is approximated as a black-body spectrum with a temperature of 5800K. The radiation power is in direct proportion to the surface area of the Sun. The luminosity at Earth is roughly consistent with the measurement achieved by American Society for Testing and Materials (ASTM) \citep{ASTM:E490}, the ASTM E-490 solar spectral irradiance is based on data from satellites, space shuttle missions, high-altitude aircraft, rocket soundings, ground-based solar telescopes, and modeled spectral irradiance,more details can be found at website \footnote{https://www.pveducation.org/pvcdrom/appendices/standard-solar-spectra}, as shown in the Figure \ref{fig2}. The peak luminosity is around 0.5 $\mu$m and the total emission power is 1366.1 $W/m^2$ at the Earth. The density of sunlight photon $n_{pn}(\nu,r)$ dependent on the photon frequency $\nu$  and the distance to the Sun $r$, which can be written in the form of:
\begin{equation}
n_{ph}(\nu,r)=\frac{2\pi \nu^2 R_{\bigodot}^2}{c^3r^2}\frac{1}{e^{h\nu/KT}-1}
\end{equation}
where $c$ is the speed of light, $R_{\bigodot}$=6.963$\times 10^{10}$ cm is the radius of  the Sun, T is the temperature, K is the Boltzmann constant.

The cross section of the $\gamma$-$\gamma$ absorption presented in  \citep{2010herb.book.....D} is adopted here:
\begin{equation}\label{eq:cross_section}
\sigma_{\gamma\gamma}(s)=\frac{1}{2}\pi r_e^2(1-\beta_{cm}^2)\left[(3-\beta_{cm}^4)ln\left(\frac{1+\beta_{cm}}{1-\beta_{cm}}\right)-2\beta_{cm}(2-\beta_{cm}^2)\right]
\end{equation}
where $r_e$ is the classical electron radius, $\beta_{cm}=\sqrt{1-s^{-1}}$, $s=\frac{1}{2}[E \epsilon(1-cos(\theta))]$ is the center-of-momentum frame Lorentz factor of the produced electron and position, $E$ is the energy of the gamma-ray, $\epsilon$ is the energy of the low energy photon, $\theta$ is the space angle between the direction of the low energy photon and that of the gamma-ray.

\begin{figure}
\begin{center}
\includegraphics[width=18cm,height=6cm]{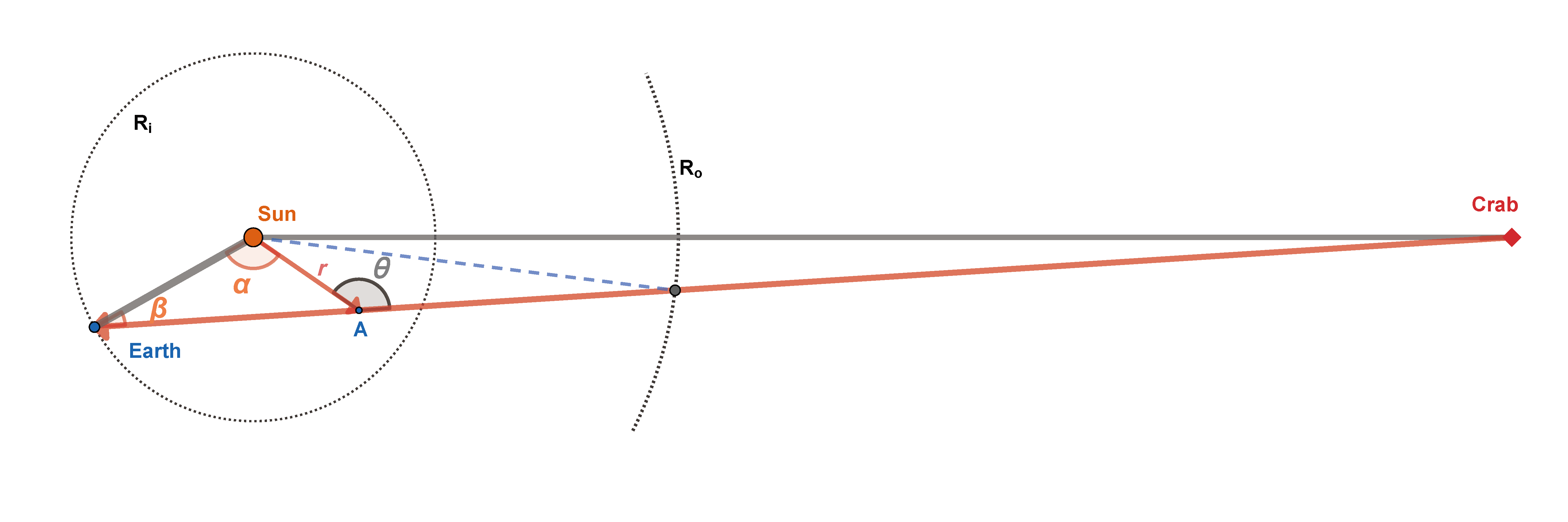}
\caption{\label{fig3} A schematic diagram shows the interaction between the gamma-rays from Crab Nebula with sunlight and the parameters used in the equations.}
\end{center}
\end{figure}

With the density of sunlight photon and the cross section of the $\gamma$-$\gamma$, the optical depth of the gamma-ray from the Crab Nebula to Earth due to the absorption of the sunlight can be estimated using the function:
\begin{equation}
\tau_{\gamma\gamma,\textit{SunLight}}= \frac{1}{2} \int dx \int_{1/\epsilon}^{\infty}  (1-cos(\theta))n_{ph}(\epsilon,r)\sigma_{\gamma\gamma}(s)  d\epsilon
\end{equation}
Where $dx$ is integrated along the line from  the Crab Nebula to the Earth as illustrated in the  schematic diagram shown in Figure \ref{fig3}. In this figure, $R_{o}$=100 AU, and the $\beta$  is the space angle between the Sun and the Crab Nebula. 
In the calculation, We divide the integration interval into two segments, i.e., with r less than $R_{o}$  and r larger than $R_{o}$. When r $<R_{o}$, the angle between the direction of the sun-light photon  and the direction the gamma-ray  change rapidly. When the  r $>R_{o}$, the collision angle will be considered as consistent value $\phi=180^{\circ}$.

\begin{figure}
\begin{center}
\includegraphics[width=13cm]{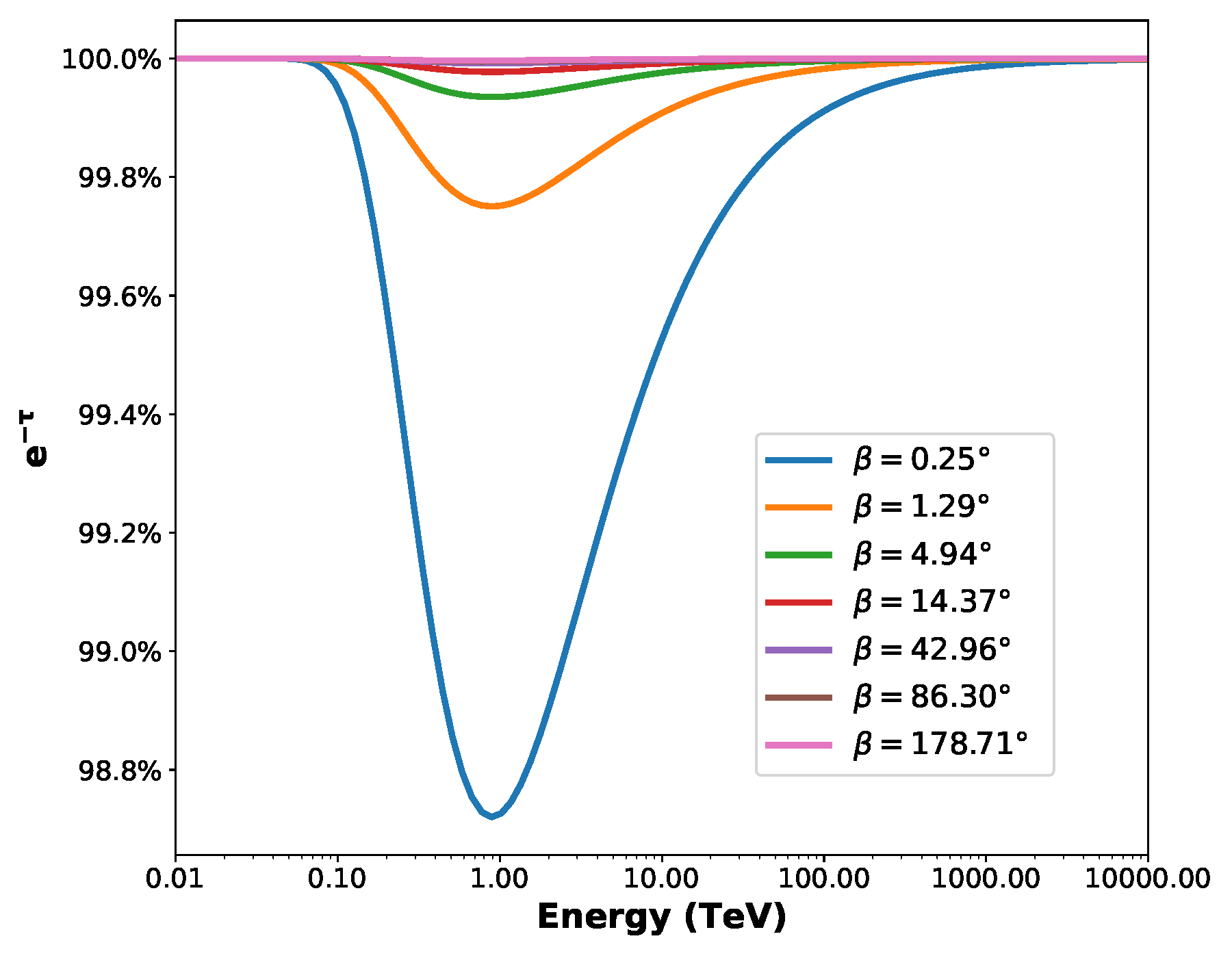}
\caption{\label{fig:angle_mjd} The survival fractions the gamma-ray after the absorption of the sunlight. Different lines indicate the results using different space angle between gamma-ray source and the Sun. }
\end{center}

\end{figure}

\subsection{Estimation result} 
Using the equation (3), we calculate the optical depth ($\tau$) of gamma-ray at different energies for different space angle between the Sun and the Crab Nebula. The survival fractions the gamma-ray $e^{-\tau}$ are shown in the Figure \ref{fig:angle_mjd}. The absorption is clearly dependent on the gamma-ray energy and the space angle between gamma-ray and the Sun. According to the Figure \ref{fig:angle_mjd}, the absorption of gamma-ray happen at a wide energy range from about 0.1 TeV to 100 TeV  with maximum absorption approximately at 0.89 TeV. When the space angle between Crab Nebula and Sun is 0.25$^{\circ}$, which is at the boundary of the Sun, the maximum absorbed percentage is 1.2\%. It is worth to note that we have ignored the absorption due to the interaction between the gamma-ray and the solar atmosphere, which should be heavy when the space angle is less than 0.25$^{\circ}$. When the space angle is 1.29$^{\circ}$, the maximum absorption ratio is 0.25\%. The maximum absorption is less than 0.05\% when the space angle is larger than 5$^{\circ}$.

\begin{figure}
\begin{center}
\includegraphics[width=13cm]{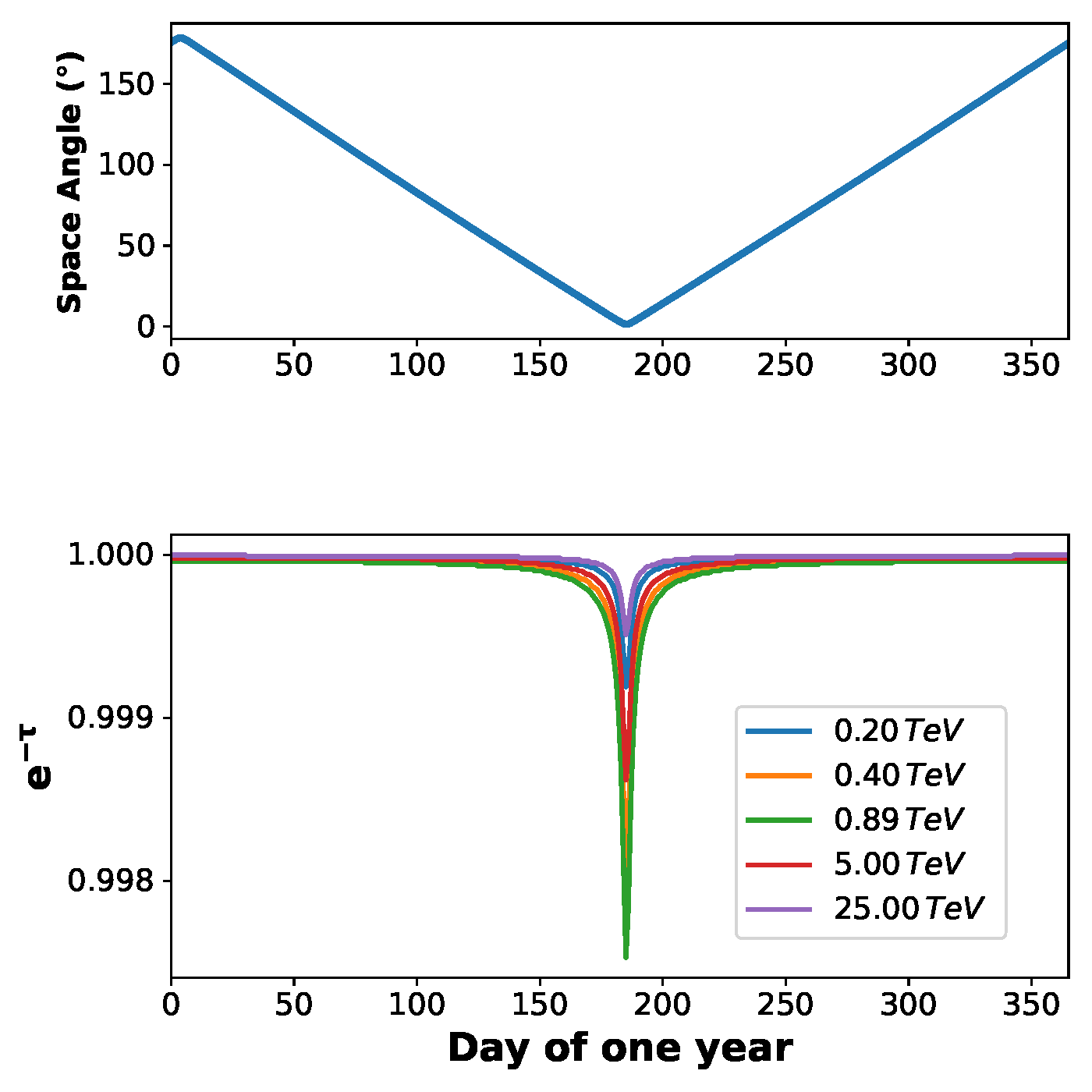}
\caption{\label{fig5} Upper: The space angle between the Sun and Crab Nebula as function of the time in one year. Down: The absorption of the sunlight on gamma-rays with different energies, i.e., 0.2, 0.4, 0.89, 5, and 25 TeV, as function of the time in one year. }
\end{center}
\end{figure}

Due to the Earth’s orbit around the Sun, the position of the Sun in the celestial coordinates varies annually. Hence the space angle between the Sun and Crab Nebula varies annually as shown in Figure \ref{fig5}, with maximum angle larger than 150$^{\circ}$ and minimum around 1.29$^{\circ}$. We also estimation of the absorption of the sunlight for gamma-rays with different energies from the Crab Nebula as the function the time as shown in Figure \ref{fig5}. Since the minimum space angle between Crab Nebula and Sun is 1.29$^{\circ}$ occurred in the day of 3th July, the maximum absorption is less than 0.25\%. 

\subsection{Cross check about the calculation procedure}
\begin{figure}
\begin{center}
\includegraphics[width=13cm]{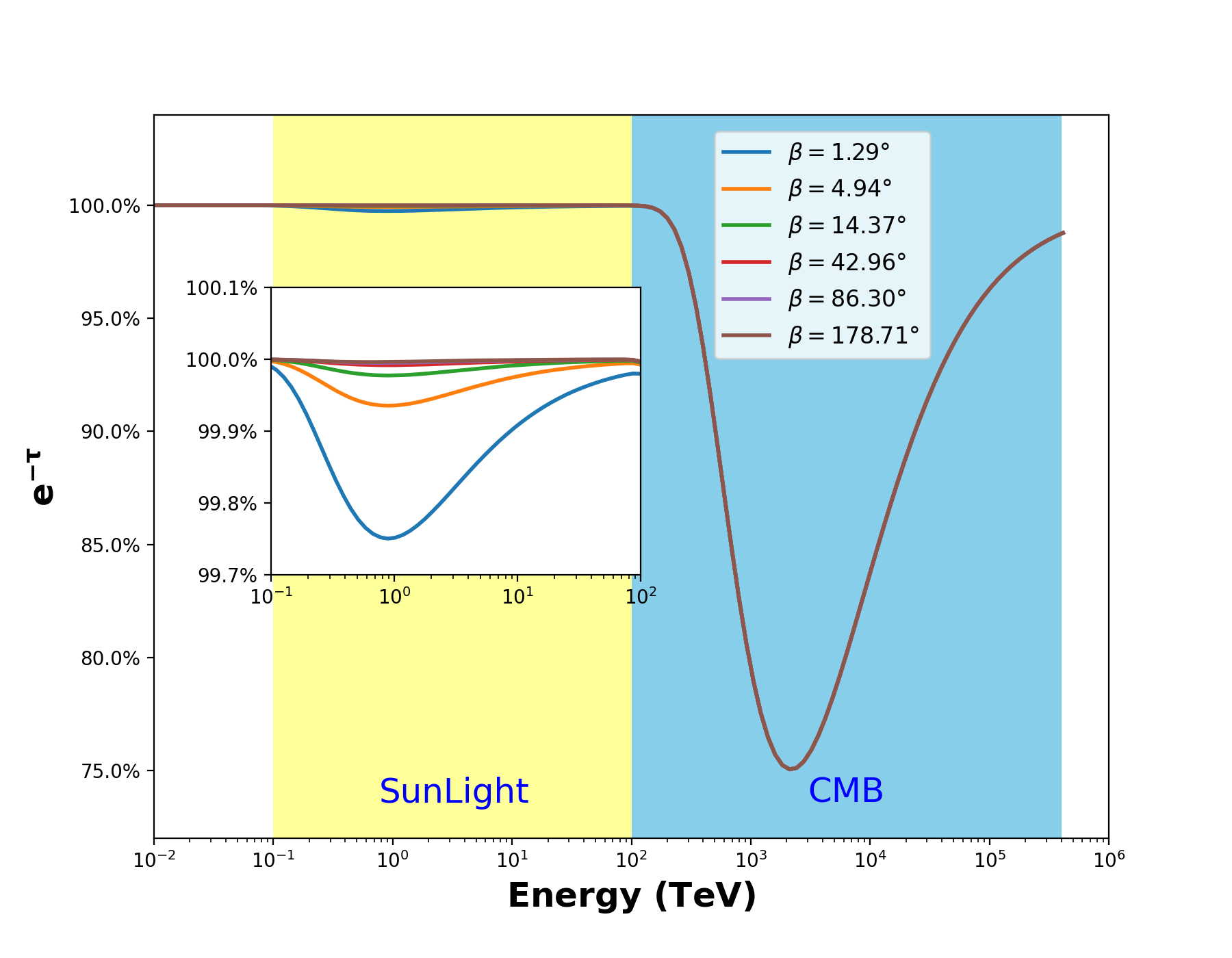}
\caption{\label{fig6} The survival fractions the gamma-ray after the absorption by the sunlight and the CMB photons. Different lines indicate the results using different space angle between gamma-ray source and the Sun.The yellow region represents the absorption of sunlight, while the sky-blue region mainly corresponds to the absorption of CMB. }
\end{center}
\end{figure}

When gamma-rays from the Crab Nebula travel towards the Earth, they are not only absorbed by low-energy sunlight photons, but also by cosmic microwave background (CMB) photons. The absorption process has been calculated in \citep{2021Natur.594...33C}. In order to validate the reliability of the calculation procedure presented in this paper, we have used the same computational procedures for sunlight photons to calculate the absorption of gamma-rays from the direction of the Crab Nebula by the CMB. In our calculations, the CMB radiation spectrum adopts a 2.7K blackbody spectrum, and the CMB density is isotropic and position-independent, in contrast to the density of sunlight photons. 
The density of CMB is:
\begin{equation}\label{eq:balck_body_cmb}
    n_{ph,\textit{cmb}}(\nu,t)=\frac{8  \pi  \nu^2}{c^3} \frac{1}{\exp\left(\frac{h \nu}{K T}\right) - 1}
\end{equation}
For the Crab, we don't need to consider the evolution of the universe, the the optical depth of CMB is:
\begin{equation}
    \tau_{\gamma\gamma,\textit{cmb}}= \frac{1}{2} \int dx \int_{-1}^{1} (1-\mu)d\mu \int_{1/\epsilon}^{\infty}  n_{cmb}(\epsilon)\sigma_{\gamma\gamma}(s)  d\epsilon
\end{equation}
where $\mu=1-cos(\theta)$.
The total survival ratio of gamma-rays as a function of energy is $e^{-\tau}=e^{-(\tau_{cmb}+\tau_{Sunlight})}$ ,
as shown in Figure \ref{fig6}. The absorption of CMB on gamma-rays  is primarily observed at energies above 200 TeV, with the maximum absorption occurring at around 2 PeV, and an absorption probability of 25\%. This result aligns with the findings in reference \citep{2021Natur.594...33C}, thus providing basic validation of the correctness of the calculations in this paper.

\section{The affection of the Sun and Moon on gamma-ray source observation}
Based on the calculations from the previous section, it is evident that the absorption of gamma-rays by sunlight is extremely minimal and can be disregarded during most observation periods. The Moon primarily reflects sunlight and has a much lower brightness compared to the Sun, resulting in an even smaller absorption of gamma-rays by moonlight, which can be directly ignored. However, when the angle between the direction of the gamma-ray source and the directions of the Sun and Moon is less than $0.25^\circ$, the gamma-rays will be completely blocked by the Sun and Moon, and cannot be ignored during this time period. Therefore, it is necessary to exclude this data during the analysis.

Additionally, when the angle between the direction of the gamma-ray source and the directions of the Sun and Moon is small, the Sun and Moon directly block the cosmic ray background, creating a missing negative source. This shadow can counteract the gamma-ray signal, and its impact depends on the ratio of the number of gamma-ray signals to the background. For previous arrays like ARGO-YBJ, which lacked the ability to distinguish between gamma-ray and cosmic ray background, the gamma-ray signals only accounted for a small portion of the backgrounds. For example, in observations of the brightest gamma-ray source, the Crab Nebula, the gamma-ray signals were only about 1\% of the backgrounds \citep{2015ApJ...798..119B}, leading to a significant counteracting effect from the shadow on the gamma-ray signals.

For the LHAASO array, due to its excellent ability to distinguish between gamma-rays and cosmic ray backgrounds, the proportion of gamma-rays to backgrounds has significantly increased. For example, in observations of the Crab Nebula, the ratio of gamma-ray signals to backgrounds  for the LHAASO-WCDA array is 20\% at 1 TeV, 80\% at 6 TeV\citep{2021ChPhC..45h5002A}, and for the LHAASO-KM2A array, the corresponding ratios are 12\% at 10 TeV, 260\% at 25 TeV, and 4100\% at 100 TeV\citep{2021ChPhC..45b5002A}. However, considering the current detection sensitivity of LHAASO is at 1\% of the Crab flux level, the impact at energies bellow 100 TeV cannot be ignored.

\begin{figure}
    \centering
    \includegraphics[width=16cm]{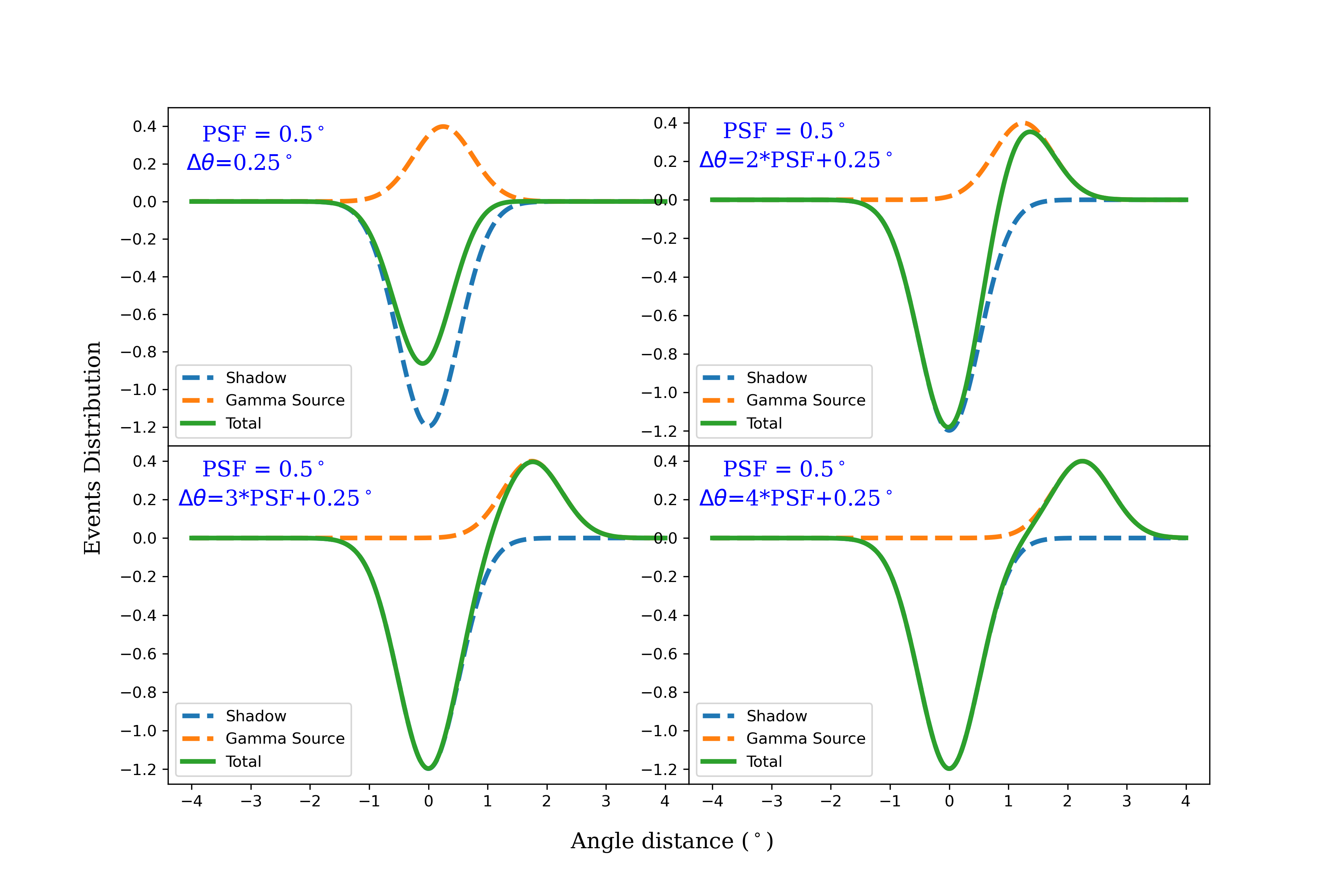}
    \caption{The impact of the shadow caused by the Sun or Moon on the observation of gamma-ray sources at different space angles, i.e., 0.25$^{\circ}$, 2*PSF+0.25$^{\circ}$, 3*PSF+0.25$^{\circ}$, and 4*PSF+0.25$^{\circ}$. The detector angular resolution (denoted as PSF) is assumed to be 0.5$^{\circ}$. The number of deficit shadow events  is assumed to be three times that of the gamma-ray signals from the source.}
    \label{fig:shadow}
\end{figure}

Fig.\ref{fig:shadow} shows the counteracting effect of Sun or Moon shadow on the gamma-ray signals at different space angles. In this example, we assume the detector angular resolution to be 0.5$^{\circ}$ and the number of deficit shadow events  to be three times that of the gamma-ray signals. This situation is similar to the  observation of LHAASO-KM2A  on a gamma-ray source with a flux around 10\% Crab unit at 10 TeV. It is clear that the gamma-ray signals will be completely counteracted by the shadow when they are very close the direction of Sun or Moon. The counteracting effect decreases as the increasing of the space angle. To ensure safety, we recommend removing data when the space angle between the gamma-ray source and the directions of the Sun and Moon is less than 3 times the angular resolution plus $0.25^\circ$. For example, the observation time of the EAS array for Crab is approximately 7 hours per day, and only during a few days the space angle between Crab and Sun or Moon is less than 2$^{\circ}$ in one year. The data from this period of observation can be excluded for data analysis, thus eliminating its impact. Since the angle between the source and the directions of the Sun and Moon varies periodically, this portion of the data accounts for a very small percentage, generally less than 1\% of the total observation time in one year, and therefore does not significantly affect the observation of the source.

Table 1 lists known gamma-ray sources with a minimum angle less than $2^\circ$ between the source and the Sun or Moon. When measuring their radiation using EAS arrays such as LHAASO and HAWC, special attention should be paid to excluding data when the space angle between the source and the Sun or Moon is small.

\section{summary}
With significant advancements in ground-based extensive air shower arrays, numerous very high-energy (VHE) gamma-ray sources have been discovered and are being continuously monitored by LHAASO and HAWC, regardless of day or night. With investigating the celestial coordinates of the Sun and Moon, we identified 20 sources that can come close to the line of sight of the Sun or Moon during some certain periods.
We have conducted the first estimation of sunlight absorption on very high-energy gamma-rays. The absorption primarily affects gamma-rays with energies ranging from 0.1 TeV to 100 TeV, with maximum absorption occurring around 0.9 TeV. The absorption is contingent on the spatial angle between the Sun and the sources; however, the overall absorption is minimal, with a maximum of less than 1.2\%. The primary influence is the obstruction of the Sun and Moon on gamma-rays, and their shadow on the cosmic ray background. The impact is also contingent on the detector's angular resolution. To mitigate its effects, we recommend excluding data when the angle between the gamma-ray source and the directions of the Sun and Moon is less than a certain value taking into account the experiment angular resolution.

\begin{acknowledgements}
This work is supported by the National Science Foundation of China under Grant Nos.12393854, 12022502, 12263007.
\end{acknowledgements}

\label{lastpage}


\begin{thebibliography}{99}

\bibitem[Aartsen et al.(2021)]{2021PhRvD.103d2005A}
Aartsen M.~G., Abbasi R., Ackermann M. et al. 2021, \prd, 103,
      042005

\bibitem[Abdo et al.(2011)]{2011ApJ...734..116A}
Abdo A.~A., Ackermann M., Ajello, M. et al. 2011, \apj, 734, 116

\bibitem[Aharonian et al.(2021{\natexlab{a}})]{2021ChPhC..45h5002A}
Aharonian, F., An, Q., Axikegu, et al. 2021{\natexlab{a}}, Chinese
  Physics C, 45, 085002

\bibitem[Aharonian et al.(2021{\natexlab{b}})]{2021ChPhC..45b5002A}
  Aharonian,F.,An,Q.,Axikegu, et al. 2021{\natexlab{b}}, Chinese Physics C, 45, 025002

\bibitem[Aielli et al.(2011)]{2011ApJ...729..113A}
Aielli, G., Bacci, C., Bartoli, B., et al. 2011, \apj, 729, 113

\bibitem[Albert et al.(2018)]{2018PhRvD..98l3011A}
Albert, A., Alfaro, R., Alvarez, C., et al. 2018, \prd, 98, 123011

\bibitem[Albert et al.(2020)]{2020ApJ...905...76A}
   Albert, A.,Alfaro, R.,Alvarez, C. et al. 2020, \apj, 905, 76

\bibitem[Albert et al.(2023)]{2023PhRvL.131e1201A}
  Albert A., Alfaro R., Alvarez C. et al. 2023, \prl, 131, 051201

\bibitem[Ambrosio et al.(1998)]{1998PhRvD..59a2003A}
Ambrosio, M., Antolini, R., Aramo, C., et al. 1998, \prd, 59, 012003

\bibitem[Ambrosio et al.(2003)]{2003APh....20..145A}
Ambrosio, M., Antolini, R., Baldini, A., et al. 2003, Astroparticle
  Physics, 20, 145

\bibitem[American Society for Testing and Materials(2000)]{ASTM:E490}
American Society for Testing and Materials. 2000, Standard Solar Constant and
  Zero Air Mass Solar Spectral Irradiance Tables, Tech. Rep. E490, ASTM
  International

\bibitem[Baixeras(2003)]{BAIXERAS2003247}
Baixeras, C. 2003, Nuclear Physics B - Proceedings Supplements, 114, 247

\bibitem[Bartoli et al.(2011)]{2011PhRvD..84b2003B}
Bartoli, B., Bernardini, P., Bi, X.~J., et al. 2011, \prd, 84, 022003

\bibitem[Bartoli et al.(2015)]{2015ApJ...798..119B}
  Bartoli,B., Bernardini,P., Bi,X.~J. et al.  2015, \apj, 798, 119

\bibitem[Bartoli et al.(2017)]{2017APh....90...20B}
    Bartoli, B. , Bernardini, P. , Bi, X., et al. 2017, Astroparticle Physics, 90, 20

\bibitem[Bartoli et al.(2019)]{2019ApJ...872..143B}
  Bartoli, B. , Bernardini, P. , Bi, et al.  2019, \apj, 872, 143

\bibitem[Cao et al.(2021)]{2021Natur.594...33C}
Cao, Z., Aharonian, F.~A., An, Q., et al. 2021, \nat, 594, 33

\bibitem[Cao et al.(2023)]{2023arXiv230517030C}
Cao, Z., Aharonian, F., An, Q., et al. 2023, arXiv e-prints,
  arXiv:2305.17030

\bibitem[Dermer \& Menon(2010)]{2010herb.book.....D}
Dermer, C.~D., \& Menon, G. 2010, {High Energy Radiation from Black Holes.
  Gamma Rays, Cosmic Rays, and Neutrinos}

\bibitem[DeYoung \& {HAWC Collaboration}(2012)]{2012NIMPA.692...72D}
DeYoung, T., \& {HAWC Collaboration}. 2012, Nuclear Instruments and Methods
  in Physics Research A, 692, 72

\bibitem[Finke et al.(2010)Finke, Razzaque, \&
  Dermer]{2010ApJ...712..238F}
Finke, J.~D., Razzaque, S., \& Dermer, C.~D. 2010, \apj, 712, 238

\bibitem[Hauser \& Dwek(2001)]{2001ARA&A..39..249H}
Hauser, M.~G., \& Dwek, E. 2001, \araa, 39, 249

\bibitem[He \& the LHAASO~Collaboration(2018)]{He2018}
He, H., \& the LHAASO~Collaboration, F. 2018, Radiation Detection Technology
  and Methods, 2, 7

\bibitem[{Jung(2001)}]{Jung2001}
Jung, I. 2001, H.E.S.S. - The High Energy Stereoscopic System, ed. M.~M.
  Shapiro, T.~Stanev, \& J.~P. Wefel (Dordrecht: Springer Netherlands),
  345--348

\bibitem[Li et al.(2024)Li, Ng, Chen, Nan, \& He]{CPC10.1088}
Li, Z., Ng, K. C.~Y., Chen, S., Nan, Y., \& He, H. 2024, Chinese Physics C, 48,
  045101

\bibitem[Linden et al.(2018)Linden, Zhou, Beacom, Peter, Ng, \&
  Tang]{2018PhRvL.121m1103L}
Linden, T., Zhou, B., Beacom, J.~F., et al. 2018, \prl, 121, 131103

\bibitem[Loeb(2022)]{2022RNAAS...6..148L}
Loeb, A. 2022, Research Notes of the American Astronomical Society, 6, 148

\bibitem[Moskalenko et al.(2006){Moskalenko}, {Porter}, \&
  Strong]{2006ApJ...640L.155M}
Moskalenko, I.~V., Porter, T.~A., \& Strong, A.~W. 2006, \apjl, 640,
  L155

\bibitem[Ng et al.(2016)Ng, Beacom, Peter, \&
  Rott]{2016PhRvD..94b3004N}
  Ng, K. C.~Y., Beacom, J.~F., Peter, A. H.~G., \& Rott, C. 2016, \prd,
  94, 023004

\bibitem[Saldana-Lopez et al.(2021)Saldana-Lopez, Dom\'\i nguez,
  {P{\'e}rez-Gonz{\'a}lez}, Finke, Ajell, {Primack}, {Paliya}, \&
  {Desai}]{2021MNRAS.507.5144S}
Saldana-Lopez, A., {Dom{\'\i}nguez}, A., {P{\'e}rez-Gonz{\'a}lez}, P.~G.,
  et al. 2021, \mnras, 507, 5144

\bibitem[Springer(2016)]{HAWC}
Springer, R. 2016, Nuclear and Particle Physics Proceedings, 279-281, 87

\bibitem[Tang et al.(2018)Tang, Ng, Linden, Zhou, Beacom, \&
  Peter]{2018PhRvD..98f3019T}
Tang, Q.-W., Ng, K. C.~Y., Linden, T., et al. 2018, \prd, 98, 063019

\bibitem[Weekes et al.(2002)]{2002APh....17..221W}
Weekes, T.~C., Badran, H., Biller, S.~D., et al. 2002, Astroparticle


\end{thebibliography}
\end{document}